\documentclass[%
 reprint,
 amsmath,amssymb,
 aps,
 pre,
]{revtex4-1}
\usepackage{graphicx}
\usepackage{grffile}
\usepackage{dcolumn}
\usepackage{bm}
\usepackage{hyperref}
\usepackage{color}

\begin{document}

\preprint{APS/123-QED}

\title{Heating leads to liquid-crystal and crystalline order in a two-temperature active fluid of rods}

\author{Jayeeta Chattopadhyay} 
\author{Sindhana Pannir-Sivajothi}
\altaffiliation[]{Department of Chemistry and Biochemistry, University of California San Diego, California 92093, USA}
\author{Kaarthik Varma}

\author{Sriram Ramaswamy}
\author{Chandan Dasgupta}
\author{Prabal K. Maiti  }
\email{maiti@iisc.ac.in}
\affiliation{%
Centre for Condensed Matter Theory, Department of Physics, Indian Institute of Science, Bangalore 560012, India
}%
\date{\today}

\newcommand{\Correction}[1]{\textcolor{black}{#1}}
\newcommand{\JC}[1]{\textcolor{red}{#1}}



\begin{abstract}

We report phase separation and liquid-crystal ordering induced by scalar activity in a system of Soft Repulsive Spherocylinders (SRS) of \Correction{shape anisotropy} $L/D = 5 $. Activity was introduced by increasing the temperature of half of the SRS (labeled \textit{`hot'}) while maintaining the temperature of the other half constant at a lower value (labeled \textit{`cold'}). The difference between the two temperatures scaled by the lower temperature provides a measure of the activity. Starting from different equilibrium initial phases, we find that activity leads to segregation of the hot and cold particles. Activity also drives the cold particles through a phase transition to a more ordered state and the hot particles to a state of less order compared to the initial equilibrium state. The cold components of a homogeneous isotropic (I) structure acquire nematic (N) and, at higher activity, crystalline (K) order. Similarly, the cold zone of a nematic initial state undergoes smectic (Sm) and crystal ordering above a critical value of activity while the hot component turns isotropic. We find that the hot particles occupy a larger volume and exert an extra kinetic pressure, confining, compressing and provoking an ordering transition of the cold-particle domains.

\end{abstract}



\maketitle

\section{\label{sec:level1} Introduction}

Active matter \cite{Sriram1, Sriram2, Romanczuk_2012} is characterized by broken detailed balance, through the conversion of a sustained supply of free energy into work at the scale of the individual constituents. This intent of this broad definition is to bring living systems into the fold of condensed-matter physics while emphasizing their nonequilibrium character. The field has advanced dramatically through experiments on scales from micrometers to kilometers highlighting the qualitative difference between active and passive systems with the same spatial symmetries, and theoretical progress uncovering the laws governing order, fluctuations, and coexistence in active systems\cite{Cates-MIPS,Ramaswamy_2003,toner2005,toner-Tu-1995,toner-Tu-1998,Hartmut2013,Hartmut2015,Hartmut2016,Hartmut2019,Hartmut2020,Marjolein-softmatter-2016,redner-baskaran,Speck2014,Golestanian-PRX-2020,Cates-PRX-2018,Stenhammar-PRL-2015,Roland2020,chaki2018entropy,chaki2019enhanced}. Simulations of minimal models are a valuable testing ground for theories and continue to present new observational puzzles \cite{Hartmut2016,redner-baskaran,Stenhammar-PRL-2015,Baskaran-softmatter,shaebani2020computational,das2017pattern}.

Heterogeneous activity is natural: motility, metabolism, or the speed of other key enzymatic processes \cite{ganai} can vary amongst the components of a system. Mixtures of motile and non-motile \cite{Cates-MIPS,Baskaran-softmatter}, or more generally active and passive, particles are another case of interest. In the simplest cases these situations are well approximated by assigning thermal baths with different temperatures to different subsets of particles \cite{ganai,joanny-2015,joanny-2018,joanny-2020}. The resulting internal heat flows make the system active in a way that is not obviously identical to the usual prescription of a maintained chemical potential difference \cite{Sriram1,Sriram2}.

Ganai et al. \cite{ganai} showed that a two-temperature picture provided a natural physical origin for chromatin organization in the nucleus, and Joanny et al. \cite{joanny-2015,joanny-2018,joanny-2020} showed analytically how phase separation arose in two-temperature systems. Spontaneous segregation in two-temperature or active-passive mixtures is widely observed in simulations, in Brownian soft disks \cite{frey}, and polymers \cite{kremer1,kremer2}, and in binary Lennard-Jones (LJ) systems \cite{Siva} where activity leads to phase separation and formation of crystalline domains. \Correction{Moreover, in literature, it is reported that non-reciprocal interactions yield two different temperatures in dusty plasmas \cite{bartnick2016emerging, PhysRevX.5.011035} and diffusiophoretic colloids \cite{PhysRevLett.112.068301}}.

These studies show the emergence of collective behavior uniquely associated with activity even when structure and dynamics at the particle scale are isotropic. Anisotropy, however, is ubiquitous in the living world in the form of the shape and movement of microorganisms, the long persistence lengths of biopolymers \cite{frey-nature,tanaka-nature} and the mesogenic nature of lipids. Liquid-crystalline (LC) order \cite{de-gennes} was central to the inception of active-matter research \cite{Sriram1}. Activity in models of liquid-crystalline order generally enters as a self-propelling force vector \cite{vicsek2012,toner2005,Sriram1,Sriram2,Chate2006,Shradha2006,Bar2006PRE,Weitz2015PRE,Bar2010PRL,huber2018,Baskaran-softmatter,Yang2010PRE,bott2018}, or an active stress tensor \cite{Simha2002,hatwalne2004,thampi2015intrinsic,santhosh2020activity}; even active isotropic baths as in \cite{AnanyoPRL2020} are created by persistent vector or tensor processes. We explore the statistical mechanics of anisotropic particles driven by a strictly scalar manifestation of activity, in a two-temperature system of soft repulsive spherocylinders (SRS). We ask:

	\begin{itemize}	
		\item How does the phase behavior of 3D soft rods depart from its equilibrium form when activity is introduced?
		\item What distinctive features can be traced specifically to the two-temperature nature of the system, in which activity leads to phase separation and resides not in any one of the particles, but at the interfaces between regions of hot and cold particles?
	\end{itemize}

In this paper, we answer these questions through molecular dynamics (MD) simulation of a collection of SRS of \Correction{shape anisotropy} $L/D = 5$ with a purely repulsive interaction. Activity is introduced by connecting half of the particles (labeled `hot') to a thermostat of higher temperature, while the rest of the particles (labeled `cold') remain connected to a thermostat of a lower temperature equal to that of the initial equilibrium system. The difference between the two temperatures scaled by the lower temperature is taken to be a measure of the strength ($\chi$) of the activity. We describe the model and the simulation protocol in detail in section II.

Our simulation study demonstrates that unlike spherical colloidal particles, where prominent effects of activity are found only when the strength of the activity is large, both for scalar ($\chi \sim 30$) \cite{joanny-2015,frey} and vector activity (Péclet number $Pe > 50$) \cite{Stenhammar-PRL-2015}, a variety of interesting phenomena are observed for colloidal rods in a much smaller range of values of the activity parameter, $1 < \chi < 4$.
This observation suggests that the two-temperature model should be an experimentally feasible system for studying the effects of scalar activity in collections of rod-like particles. In this regard, the critical activity $\chi_{c}$, defined as the value of $\chi$ at which macroscopic phase separation starts to occur, shows a non-monotonic dependence on the packing fraction $\eta$, decreasing with increasing $\eta$ in the liquid regime and increasing again in the crystalline regime. A detailed analysis of the phase separation can be found in section III-A.\\

The segregated zones develop different liquid-crystalline (LC) structures depending on the level of activity and the reference equilibrium phase of the system at zero activity. We observe the cold particles undergoing a phase transition towards a more ordered state, and the hot particles towards a less ordered state, as compared to the initial equilibrium state. If the system at zero activity is in the isotropic (I) fluid phase, the cold domains that emerge are nematic (N) and, at higher activities, crystalline (K), while the hot particles remain in the isotropic phase with reduced density. Similarly, a homogeneous nematic reference configuration shows smectic (Sm) and crystalline cold domains and isotropic structure in the hot domain. As a result, the phase boundary of the I-N transition shifts towards lower density for the cold particles and higher density for the hot particles. Different LC phases are identified by calculating the local nematic order parameter and suitable pair correlation functions. 
 Finally, we analyse interfacial properties and find that LC ordering in the lower packing fractions (starting from I, N phase) is governed by local balance of pressure across the interface: higher temperature induces higher kinetic pressure in the hot zone which is compensated in the cold zone by increasing virial pressure. Thus mechanical stability is maintained at the interface. Detailed analyses of the segregated phases and interfacial properties are presented in sections III-B,C,D.\\

The extraordinary nonequilibrium feature that we wish to highlight is that an enhancement of the temperature of a fraction of the particles gives rise to enhanced LC ordering in the remaining particles at any packing fraction.



\section{\label{sec:level2} Model and Simulation Details}

\begin{figure} [!htb]
	\centering
	\includegraphics[scale=0.45]{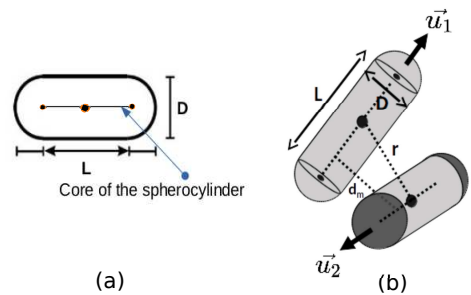}
	\caption{Schematic diagram of SRS. (a) The line segment joining the centers of the two hemispheres is known as core of the spherocylinder. (b) $\vec{u_{1}} $ and $ \vec{u_{2}} $ describe the orientations of the spherocylinder 1 and 2 respectively and $ r $ is the distance between their centers of masses. $d_{m}$ is the shortest distance that determines the interaction potential between them. \Correction{Panel (b) is adapted from reference \cite{cuetos-2002}}
	} \label{pic1}
	
\end{figure}

We model the system as a collection of SRS (cylinders with hemispherical caps). The shape anisotropy \Correction{(A)} is defined by the \Correction{ratio of length (L) and diameter (D)} $ A = L/D $ (Fig. \ref{pic1}). Spherocylinders interact through the Weeks-Chandler-Andersen (WCA) potential \cite{weeks1971} generalized to non-spherical bodies:

\begin{equation} \label{usrs}
\begin{split}
U_{SRS} & =  4\varepsilon\left[\left(\dfrac{D}{d_{m}}\right)^{12}-\left(\dfrac{D}{d_{m}}\right)^{6}\right]+\epsilon \qquad \rm{if} \quad d_{m} < 2^{\frac{1}{6}}D\\
& = 0 \qquad \qquad \qquad \qquad \qquad \qquad \qquad \rm{if} \quad d_{m} \geq 2^{\frac{1}{6}}D
\end{split}
\end{equation}
where $d_m$ is the shortest distance between two spherocylinders that determines their relative orientation and interacting force \Correction{\cite{allen1993hard,vega1994fast,earl2001}}. \Correction{Note that representing spherocylinder by a line of interacting spheres can also be used to study various LC phases \cite{HeyesPRE}}

We perform molecular dynamics (MD) simulations in the constant number-pressure-temperature (NPT) ensemble, using a Verlet algorithm \cite{verlet1967} to update the positions and velocities of the particles and quaternion-based rigid-body dynamics \cite{omelyan1998,martys1999,rotunno2004,maiti2002,lansac2003} for rotational motion. The temperature and pressure of the system are maintained using a Berendsen thermostat and manostat \cite{berendsen1984} with a temperature relaxation time $\tau_T=0.05$ and pressure relaxation time $\tau_P=2.00$ respectively. Thermodynamic and structural quantities are scaled by system parameters (i.e $\epsilon, D$) and calculated in reduced units: temperature $T^*={k_{B}T}/{\epsilon}$, pressure $P^*= \dfrac{Pv_{hsc}}{k_{B}T}$, packing fraction $\eta = v_{hsc}\rho$, where $\rho = \frac{N}{V}$ and $v_{hsc}= \pi D^{2}(\frac{D}{6} + \frac{L}{4})$ is the volume of the spherocylinder.\\
 
We prepare the system initially in a hexagonal close packed (HCP) crystalline structure. As the constituent particles are asymmetrical in shape, we choose the numbers $n_x$, $n_y$, $n_z$ of unit cells in the $x$, $y$, and $z$ directions respectively in such a way that the simulation box can be constructed in a nearly cubic geometry. If $ n_u $ is the number of spherocylinders in one unit cell then the total number of spherocylinders $N = n_u \times n_x \times n_y \times n_z$. The usual periodic boundary condition and minimum image condition are used. A system of $N = 1024$ is built by choosing $n_x = n_y = 16, n_z = 4 $ . Ratios of the dimensions of simulation box are: $L_x / L_y = 1.16 ,  L_z / L_y = 1.68$. Later we increase the system size to $N = 4096$ to check for finite size effects.\\
 
After building the system, we equilibrate it at $T^*=5.00$. We then establish the equilibrium phase diagram for this temperature by slowly varying the pressure to melt the system. We simulate for a range of pressures $P^*$ from $20$ to $0.05$ which spans crystal to isotropic phases. The ordering transitions are located by calculating the nematic order parameter and suitable pair correlation functions. The order parameter for the nematic phase is a traceless symmetric tensor $ \boldsymbol{Q} $, defined below, which is used to obtain the scalar nematic order parameter $S$, which is the largest eigenvalue of $ \boldsymbol{Q} $, and the corresponding eigendirection, which is the director ${\bf n}$. A value of $S$ consistent with $0$ defines the isotropic phase. In highly ordered states, $S \simeq 1$. Let $ u_{i}^{\alpha}$ be the $\alpha^{th}$ component of the orientation vector of spherocylinder $i$. Then we define
\begin{center}
	 	$ Q_{\alpha\beta} = \frac{1}{N}\sum_{i=1}^{N}\left(\dfrac{3}{2}u_{i}^{\alpha}u_{i}^{\beta}-\dfrac{1}{2}\delta_{\alpha\beta}\right)$ 
\end{center}

We introduce activity by choosing half of the particles randomly and assigning a higher temperature to them while keeping the other particles' temperature fixed at a lower value equal to that of the initial equilibrium system. Let $ T_{h}^{*} $ and $ T_{c}^{*} $ be the temperature of the hot and cold particles respectively. Initially we equilibrate the system at  $T_{h}^{*} = T_{c}^{*} = 5.00 $, then increase $T_h^*$ in steps: 
$T_{h}^{*} = 5.00 \rightarrow 7.50 \rightarrow 10.00 .... 30.00 \rightarrow 50.00$, allowing the system to reach a steady state after each increase in $ T^*_h $, keeping the \textit{volume} of the simulation box constant throughout the simulation.  
As a result of heat exchange, the measured of effective temperatures $T_{h}^{eff}$, $T_{c}^{eff}$ of the two populations, as defined by their steady-state average kinetic energies, differ from those of their thermostats:  

\begin{equation}
T_{h}^{*} > T_{h}^{eff} > T_{c}^{eff} > T_{c}^{*}.
\label{eq2}
\end{equation} 

We parameterize activity by 

\begin{equation}
\chi = \dfrac{ T_{h}^{*} -  T_{c}^{*}}{ T_{c}^{*}}
\end{equation}

For the active case, i.e., for $\chi \neq 0$, we choose the thermostat relaxation time $\tau_{T} = 0.01$ 
for both types of particles. We run the simulation for $3\times10^{5}$ to $4\times10^{5}$ integration time steps to reach steady state and another $10^{5}$ steps to calculate thermodynamic and structural quantities. We use an integration time-step $\delta t = 0.001$ in units of the natural timescale $D\sqrt{m/\epsilon}$.



\section{\label{sec:level3} Results and analysis}

\begin{figure*} [!htb]
	\centering
	\includegraphics[scale=0.80]{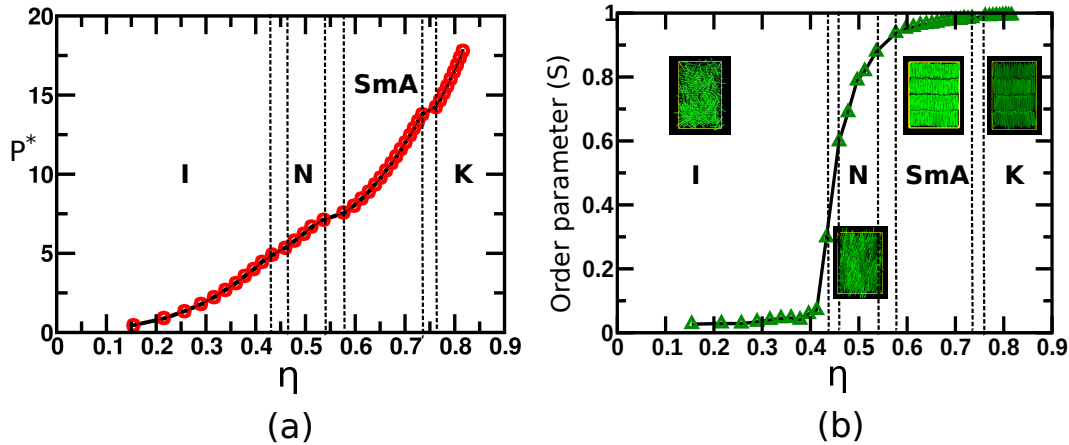}
	\caption{(a) The equation of state and (b) nematic order parameter $S$ vs packing fraction $\eta$ for the system of soft repulsive spherocylinders with \Correction{shape anisotropy} $L/D = 5.00$. Thermodynamic quantities are defined in reduced unit: packing fraction $ \eta = \rho v_{hsc} $ and pressure $P^*= Pv_{hsc}/kT$ where $ v_{hsc} $ is the volume of the spherocylinder. Here we observe four stable phases: isotropic (I), nematic (N), smectic (Sm) and crystal (K). The vertical dashed lines indicate coexisting regions near the phase transition points. } \label{eos-op-5}
	
\end{figure*}

We present the equilibrium phase diagram of SRS for $L/D = 5 $  at $T^{*} = 5 $ and observe four stable phases: (i) crystal (K), (ii) smectic A (SmA), (iii) nematic (N), (iv) isotropic (I) (Fig. \ref{eos-op-5}). The critical values of thermodynamic quantities at phase transition points match well with previous results by Cuetos et al. \cite{cuetos-2002,cuetos-2015}.

\subsection{Activity-induced phase separation} 

\begin{figure*} [!htb]
	\centering
	\includegraphics[scale=0.80]{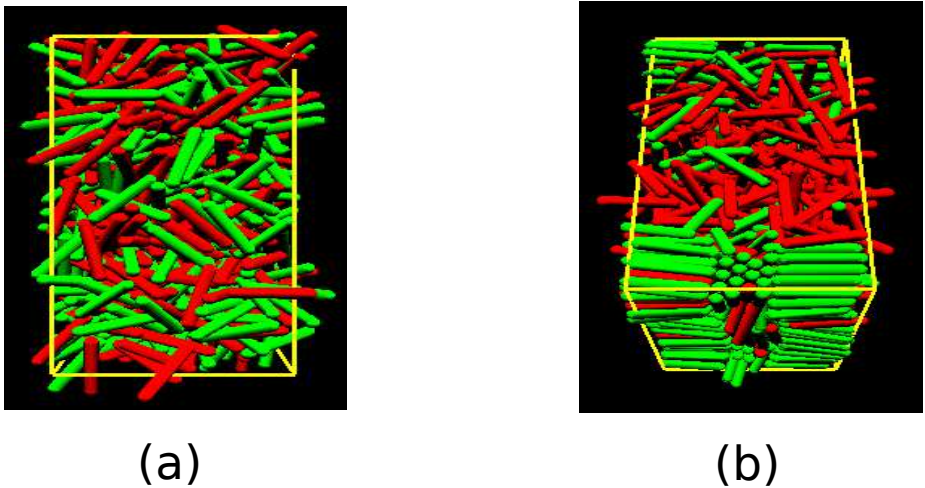}
	\caption{(a) Equilibrium configuration of N = 1024 soft spherocylinders at the state point $\eta = \rho v_{0} = 0.36, T^{*} = 5.00 $ in the absence of activity $ \chi = 0.00 $. Both hot (red) and cold (green) particles are well mixed at the same temperature. (b) Steady state configuration after phase separation at $ \chi = 5.00 $. \Correction{ It is clearly visible that cold particles are segregated and ordered, whereas the surrounding hot particles are disordered.}} \label{phase-sep-5}

\end{figure*}

Our system receives a sustained flux of energy which the hot particles draw from the hot bath and transfer through collisions to the cold particles which in turn reject the excess to the cold bath. At a steady state, the power gained by the hot particles is equal to the power transferred by the cold particles keeping the net energy flux into the system zero. A region occupied predominantly by hot particles tends to expand relative to one with cold particles. This opens up the possibility of phase separation by hot particles self-consistently sequestering a domain of cold particles. As $\chi$ is increased, we indeed see such an effect, locally at first and then macroscopically with a well-defined interface (Fig. \ref{phase-sep-5}). \\

\Correction{The extent of phase separation is quantified from the spatial distribution of hot and cold particles. To do so, we divide the simulation box into a number of sub-boxes ($ N_{box} $) and for each sub-box ($ i $) , we calculate the absolute number difference of hot ($ n_{h}^{i} $) and cold ($ n_{c}^{i} $) particles  divided by total number of particles in that sub-box. This quantity is denoted as order parameter and is averaged over all the sub-boxes and also over sufficiently large number of steady state configurations as given by the following equation:} 

\Correction{
\begin{equation}
\phi=\dfrac{1}{N_{box}} \left\langle\sum_{i=1}^{N_{box}}\frac{\lvert n^{i}_{c}-n^{i}_{h}\rvert}{(n^{i}_{c}+n^{i}_{h})} \right\rangle _{ss} \label{phase-sep-eq}
\end{equation} }

\Correction{where $\langle...\rangle_{ss}$ denotes 
a steady state average over a sufficiently large number of configurations. The selection of number of sub-boxes is arbitrary; we choose it such that (in our case, $ N_{box} = 4^{3} $) each box contains enough particles to obtain good statistics.
Ideally, in the absence of activity (at $ T_{c}^{*} = T_{h}^{*} = 5.00 $), $ \phi $ should be zero. But for a finite system size, it can be non-zero, hence we offset it by the initial value ($\phi_{0} $ ), $\phi \to \phi - \phi_{0} $. }\\

\begin{figure} [!htb]
	\centering
	\includegraphics[scale=1]{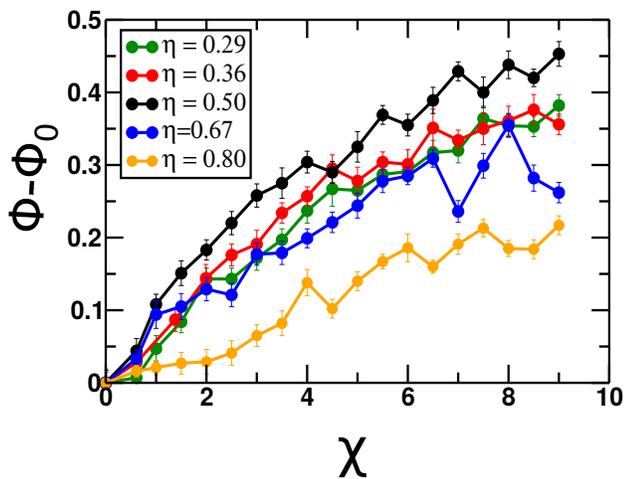}
	\caption{\Correction{Density order parameter $ \phi $ vs activity $ \chi $ at several packing fractions ($ \eta $) of the system.}
	} \label{denop}
	
\end{figure}

In Fig. \ref{denop}, we observe $\phi$ increases monotonically with $\chi$ up to a certain value, then saturates. The reason is, local separation emerges at lower activities which increases until a well-defined interface is formed (\Correction{See Appendix for detailed calculation of macroscopic phase separation.}).  
The value of $\chi$ at which phase separation starts to occur macroscopically is defined as the critical activity $\chi_{c}$. But calculating $ \chi_{c} $ from Fig. \ref{denop} is difficult as the crossover between mixed and phase-separated states is not sharp enough. Hence, we identify $\chi_{c}$ from the following criteria: 
we define a quantity $\psi$ that signifies the number difference between hot and cold particles in each sub-box: $ \psi = \langle \frac{n_{c}-n_{h}}{n_{c}+n_{h}} \rangle _{ss} $ and compute the distribution $P(\psi)$ over the sub-boxes. The activity at which $P(\psi)$ develops bimodality is considered to be the critical activity $\chi_{c}$ of the system.

\begin{figure*} [!htb]
	\centering
	\includegraphics[scale=1.3]{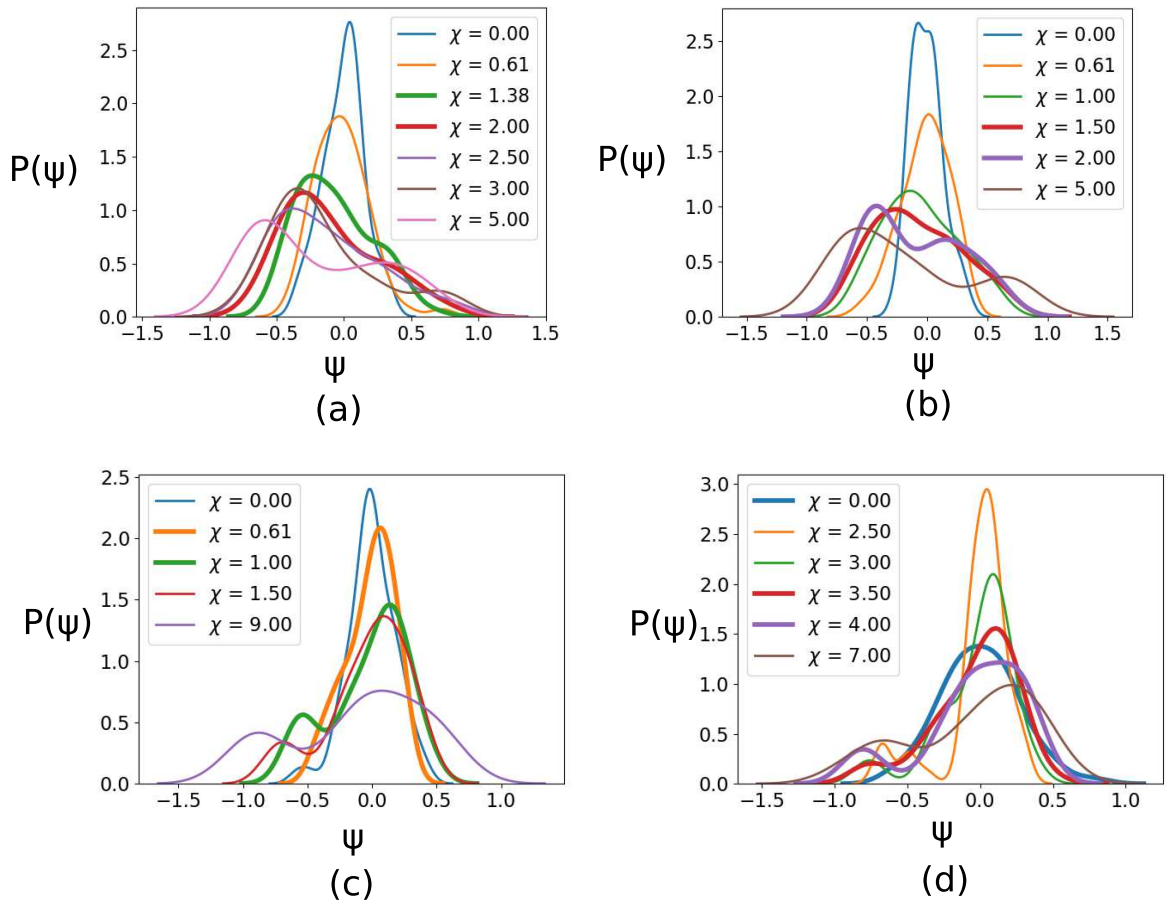}
	\caption{Distribution of $ \psi $, $ P(\psi) $, at different activities $ \chi $ for several different packing fractions $ \eta $. Critical activity $ \chi_{c} $ is defined as the value of $ \chi $ at which $ P(\psi) $ develops bimodality. However, we find a plateau region just below the calculated $ \chi_{c} $ which is the signature of emergence of phase-separation. Therefore, the exact value of $ \chi_{c} $ lies within the following range: (a) $\eta = 0.36$, initial isotropic phase: $ \chi_{c} = 1.38-2.0 $ (b) $\eta = 0.50$, initial nematic phase: $ \chi_{c} = 1.50-2.00 $ (c) $\eta = 0.67$, initial smectic phase: $ \chi_{c} = 0.61-1.00 $ (d) $\eta = 0.80$, initial crystal phase : $ \chi_{c} = 3.50-4.00 $. 
	\Correction{Note that, in the crystal phase, bi-modality appears from $\chi = 2.50-3.00 $. However, we do not considered this the critical activity $\chi_{c}$ as the higher peak arises at $\psi = 0 $ indicating most of the particles are mixed. In contrast, at $\chi = 3.50-4 $, both of the peaks occur at a nonzero value of $\psi$ ($\psi = -0.8, 0.25$).}
	$ P(\psi) $s for the two limits on $ \chi_{c} $ are shown as thick lines. The range of $ \chi_{c} $ is represented by a gray color band in the phase diagram, Fig. \ref{crit-chi-eta}. \Correction{The non-zero weight for 
	$|\psi| > 0 $ arises from the fitting procedure.}} \label{crit-denop}
	
\end{figure*}

\begin{figure} [!htb]
	\centering
	\includegraphics[scale=0.80]{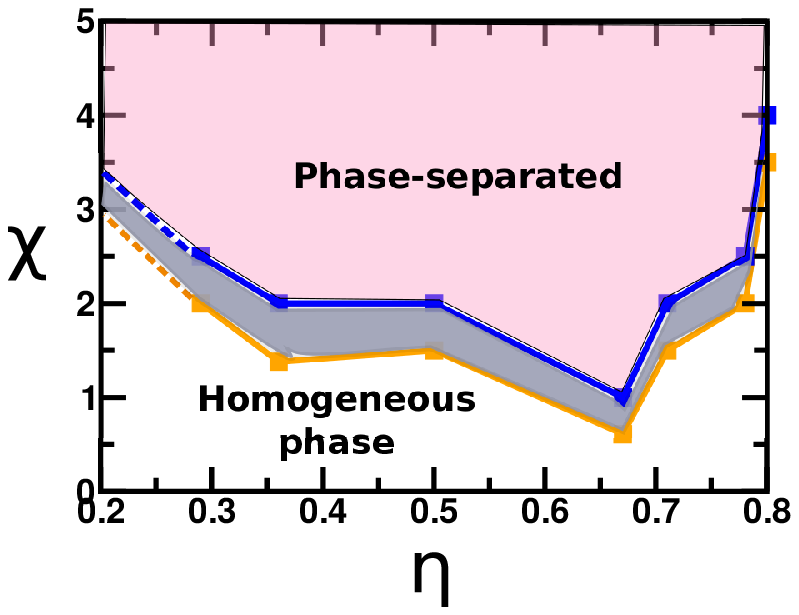}
	\caption{Phase Diagram in the state space  $ \chi$ vs $\eta $. The pink-shaded area indicates the phase-separated region and the non-shaded area indicates the region of the homogeneous phase where hot and cold particles are well mixed. The blue and orange lines indicate upper and lower limits of critical activity $ \chi_{c} $ and the in-between area of gray shade  denotes the range of possible values of $ \chi_{c} $. Dotted lines are extrapolations from the calculated data.
	} \label{crit-chi-eta}

\end{figure}

In Fig. \ref{crit-denop}, we calculate $ \chi_{c} $ from $ P(\psi) $ for different packing fractions corresponding to the different initial phases. In the case of lower packing fractions ($ \eta = 0.36 $, Fig. \ref{crit-denop}-a), bimodality appears at a higher value than actual $ \chi_{c} $. However, we observe a plateau regime with shifted unimodal peak which is the signature of emergence of phase separation. This is also seen for other packing fractions just below the calculated $ \chi_{c} $. Therefore, for each $ \eta $, we define a range of $ \chi $ within which the exact value of $ \chi_{c} $ lies. With these observations, we present a complete phase diagram in the state space,  $( \chi$ vs $\eta )$, showing parametric regions of mixed and phase separated states (Fig. \ref{crit-chi-eta}).

From Fig. \ref{crit-chi-eta}, we find that $ \chi_{c} $ decreases with the increase of packing fraction $ \eta $ up to a value of $\eta = 0.67 $. This can be due to the fact that the interaction between hot and cold particles is higher for dense systems which causes fast dissipation of hot particles' energy. Beyond  $ \eta = 0.67 $, crystalline order emerges and $ \chi_{c} $ increases again as a function of $ \eta $ (Fig. \ref{crit-denop}-d). The possible reasons are:(i) in extremely dense system a lot of hot particles are stuck in cold zone which require a larger amount of energy to overcome the barrier (ii) the relaxation is very slow in the crystal phase compared to the liquid crystal phases. Therefore, it may require a longer time to undergo phase separation at smaller activities. However, it is interesting to note that critical activity lies in a very small range $ 1.0 < \chi_{c} < 4.0 $ i.e ratio of temperatures $ 2.0 < T_{h}^{eff}/T_{c}^{eff} < 5.0 $ for the entire range of $\eta $. This observation indicates that two-temperature model should be a reliable system to observe the effect of scalar activity in colloidal rods experimentally.

\subsection{Activity-induced liquid-crystalline ordering}

Hot particles exert an active kinetic pressure along the hot-cold interfaces, which drives an ordering transition in the cold particles. The ordered structures in the phase-separated domains depend on the overall packing fraction $\eta$, $T^*$, and $\chi$. Starting from the state points in the equilibrium $\eta$-$P^*$ phase diagram corresponding to isotropic, nematic, and other phases, we observe the cold domains undergoing phase transitions towards more ordered states and the hot domains towards less ordered states, as compared to the initial equilibrium state. The extent of the segregated zone is quantified by the density profile normal to the interface which we discuss later (in section III-D). Different phases are characterized by calculating the local nematic order parameter $S$ and suitable positional and orientational pair correlation functions.

\subsubsection{Initial Isotropic configuration}

\begin{figure*} [!htb]
	\centering
	\includegraphics[scale=0.90]{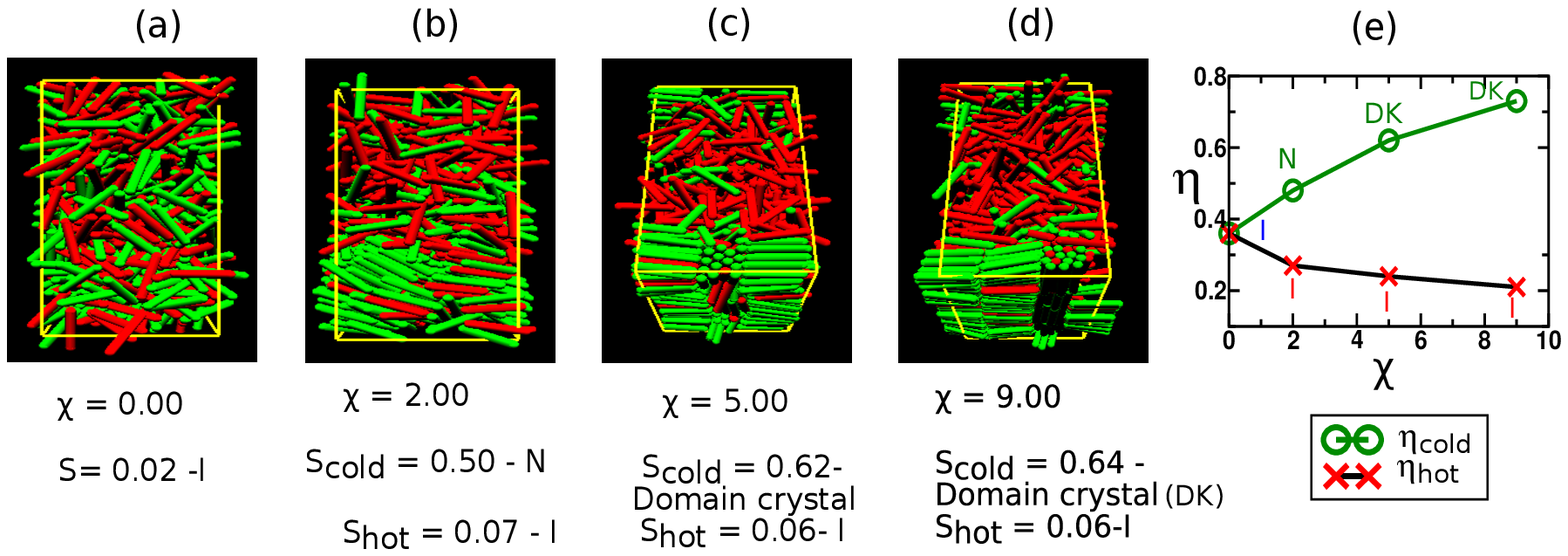}
	\caption{Segregated structures for initial isotropic phase (I) $\eta = 0.36$: (a) initial configuration in the absence of activity (b) nematic (N) ordering in the cold cluster; (c, d) multi-domain crystal (DK) in the cold cluster. \Correction{(e) Packing fractions in the segregated zone corresponding to the aforementioned activities. The phases in each zone are mentioned for each activity. $S_{cold}$ and $S_{hot}$ are the nematic order parameters of cold and hot particles respectively. $\eta_{cold}$ and $\eta_{hot}$ are the packing fractions in the cold and hot zone respectively.} In the case of (c, d), $S_{cold}$ is much lower than that for usual crystalline ordering as it is calculated by averaging over all the domains with different orientations of the directors.  However, the local ordering in each domain is much higher ($S_{cold}^{loc} = 0.90 $) which indicates crystalline order. The lines drawn in Fig. (e) are guide to the eye. } \label{iso}
	
\end{figure*}

\begin{figure*} [!htb]
	\centering
	\includegraphics[scale=1.00]{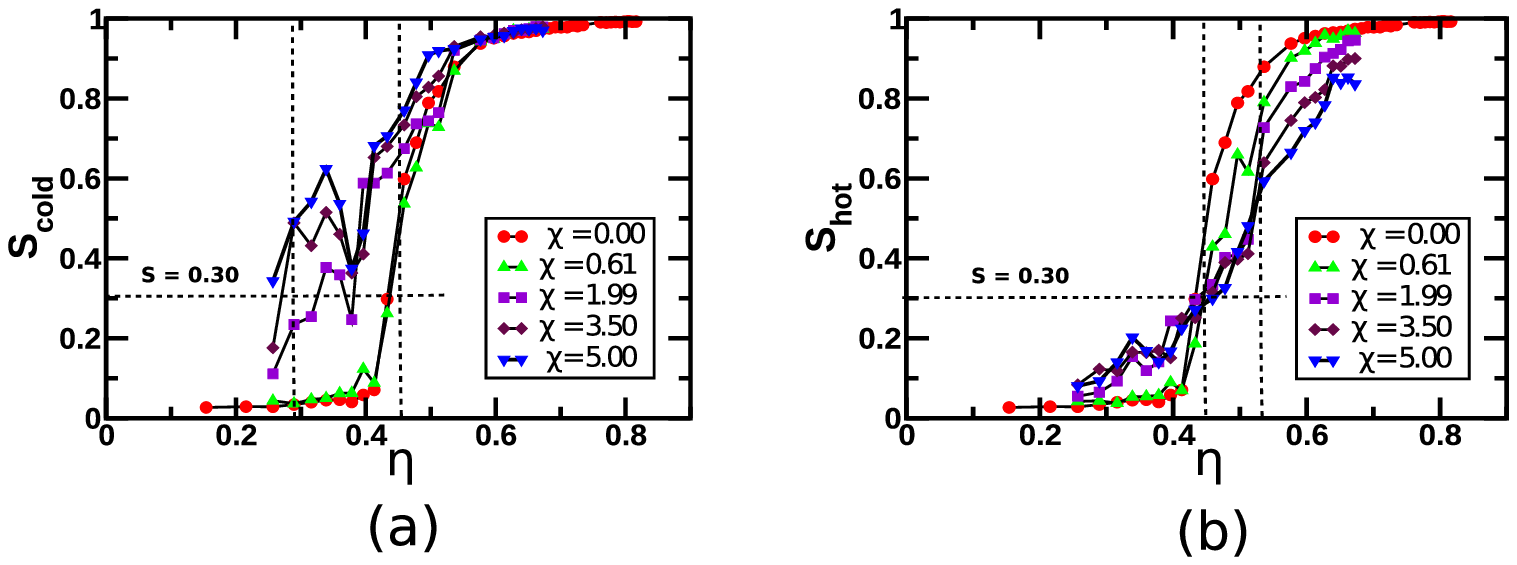}
	\caption{(a) Nematic order parameter of cold particles $ S_{cold} $ and (b) hot particles $ S_{hot} $ vs packing fraction $ \eta $ at different activities $\chi$. The vertical dashed lines indicate the shift of the I-N phase boundary towards lower packing fraction for cold particles and higher packing fraction for hot particles. The horizontal dashed line indicates the critical value of order parameter ($ S = 0.30 $) assumed to indicate the isotropic to nematic transition.
	} \label{hot-cold-5}
	
\end{figure*}

In Fig.-\ref{iso}, we show the emergence of various phases in the hot and cold regions under different activities, starting from an initial isotropic (I) phase. The critical activity $ \chi_{c} $ for phase separation lies between 1.38 to 2.00. Cold particles undergo a transition to a  nematic (N) phase at lower activities (Fig. \ref{iso}-b) which eventually turns into crystalline order at higher activities (Fig. \ref{iso}-c,d). However, hot particles remain in the isotropic phase with reduced packing fraction. Hence, the I-N phase boundary shifts towards lower density for the cold particles and higher density for the hot particles. In Fig. \ref{hot-cold-5}-(a) or (b), we see a continuous phase transition from disordered to ordered state for active systems, as is evident from the continuous increase in the nematic order parameter in contrast to the sudden jump in the order parameter for the equilibrium case. We notice a local minimum in $ S_{cold} $ between  $ \eta = 0.36-0.38$ (Fig. \ref{hot-cold-5}-a) . To check possible effects of finite system size, we simulated a larger system with $N = 4096$ SRS and observed similar results (see Appedix Fig. \ref{A-2} for details). The local minimum occurs due to the formation of multiple domains with different orientations of the nematic director, which effectively reduces the global nematic order parameter of the cold particles. This is also verified by calculating orientational and positional pair distribution functions in the cold regions of the respective densities (Fig. \ref{A-3}). For other densities, we observe a single domain with a fixed orientation of the director in the cold zone which increases $ S_{cold} $.\\

\subsubsection{Initial Nematic configuration}

\begin{figure*} [!htb]
	\centering
	\includegraphics[scale=0.90]{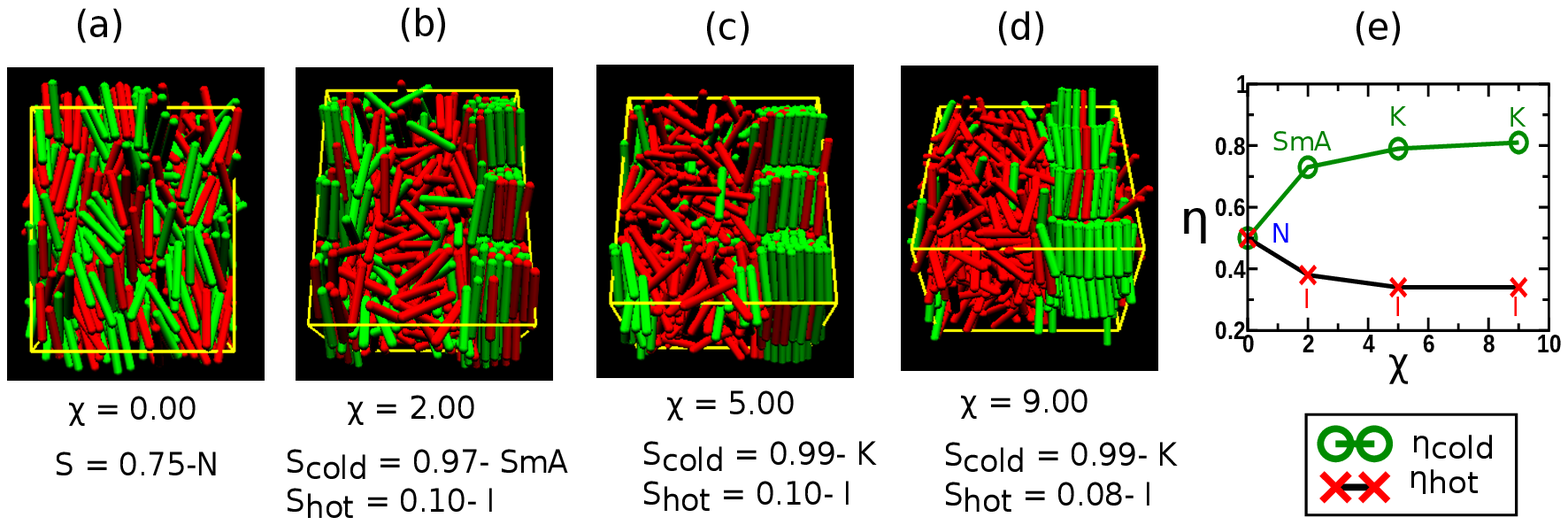}
	\caption{Segregated structures in a nematic (N) initial configuration ($ \eta = 0.50 $):
	(a) initial structure in the absence of activity. (b) smectic (Sm) ordering in the cold zone and isotropic (I) structure in the hot zone;
	(c,d) crystalline (K) ordering in the cold zone and isotropic structure with reduced density in the hot zone. 
	\Correction{(e) Packing fractions in the segregated zone corresponding to the aforementioned activities. The phases in each zone are mentioned for each activity. Parameters are the same as mentioned in Fig. \ref{iso}.}} \label{nem}
	
\end{figure*}

For the initial nematic configuration, phase separation starts at $ \chi_{c} = 1.50-2.00 $. We found that activity drives the cold particles to undergo a nematic to smectic (N-Sm) transition while the hot particles exhibit a nematic to isotropic (N-I) transition as shown in Fig. \ref{nem}. In Fig. \ref{nem}-b, we can see that at $ \chi = 2.00 $, the nematic order parameter in the cold zone is $ S_{cold} = 0.97 $, and the packing fraction is $ \eta_{cold} = 0.73 $. These values are consistent with the equilibrium smectic phase for SRS with a \Correction{shape anisotrpy} $L/D = 5$ (Fig. \ref{eos-op-5}). On the other hand, the hot particles develop isotropic structure with $ S_{hot} = 0.10 $ and $ \eta = 0.38 $. Further increase of $ \chi $ turns the cold zone into a close packed crystal structure as shown in Fig. \ref{nem}-(c), (d).

\subsubsection{Initial Smectic configuration}

\begin{figure*} [!htb]
	\centering
	\includegraphics[scale=0.90]{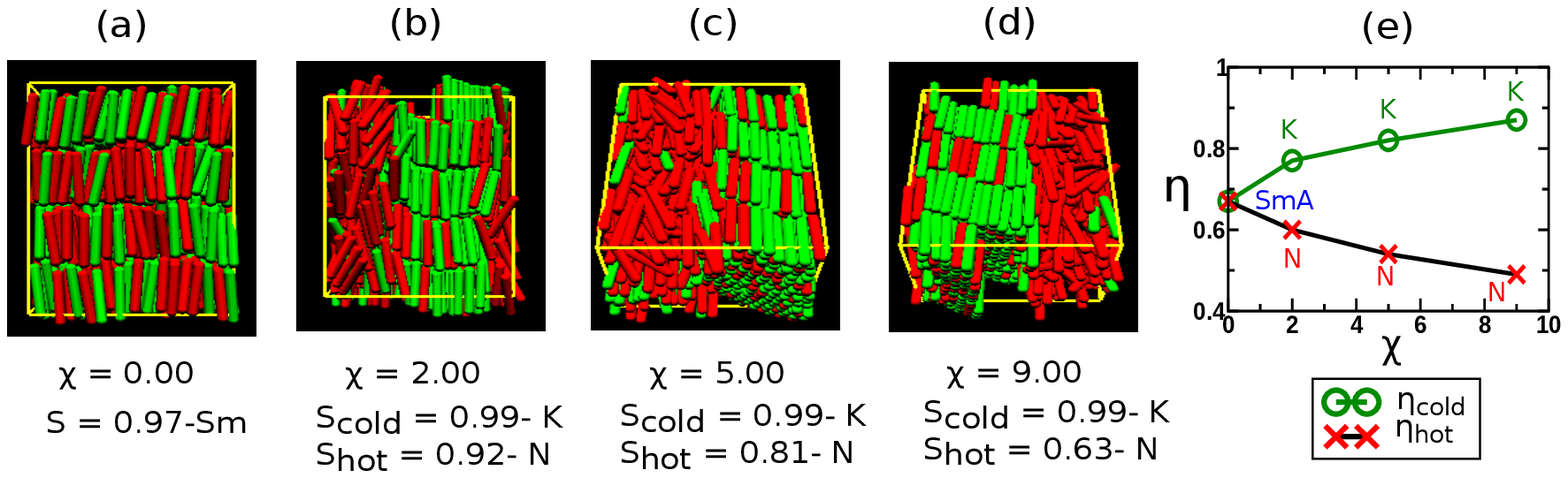}
	\caption{Segregated structures in a smectic (Sm) initial phase ($ \eta = 0.67 $): (a) initial configuration in the absence of activity (b\textemdash d) crystal (K) structures in the cold region and nematic (N) structures in the hot region at different activities.
	\Correction{(e) Packing fractions in the segregated zone corresponding to the aforementioned activities. The phases in each zone are mentioned for each activity. Parameters are the same as mentioned in Fig. \ref{iso}.}} \label{sm}
	
\end{figure*}

In the case of an initial smectic configuration, the system starts to phase separate at a very low activity: $ \chi_{c} = 0.61-1.00 $. A small amount of temperature difference drives the cold zone into a close packed crystal structure while the hot zone undergoes a transition to the nematic phase, as shown in Fig. \ref{sm}. For $\chi = 9.0 $, the local nematic order parameters and packing fractions in the segregated regions are: hot region: $ S = 0.58 $, $ \eta = 0.50 $ which is consistent with the equilibrium nematic phase; cold region: $ S = 0.98 $, $ \eta = 0.87 $ which is consistent with the equilibrium crystal phase (Fig. \ref{eos-op-5}). The observed phases are further verified by calculating suitable pair correlation functions as discussed in detail in section III-C.

\subsubsection{Initial Crystal configuration}

\begin{figure*} [!htb]
	\centering
	\includegraphics[scale=0.90]{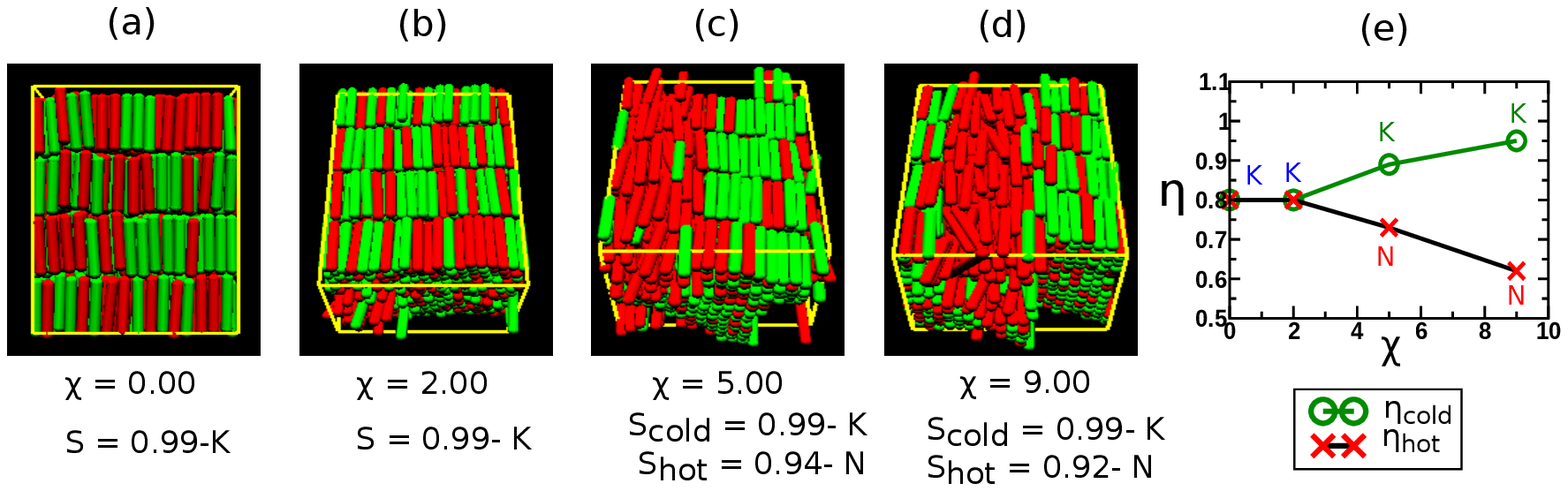}
	\caption{Segregated structures in a crystalline (K) initial state ($ \eta = 0.80 $):
		(a) initial configuration in the absence of activity (b) no phase-separation at $ \chi = 2.00 $, both hot and cold particles are in crystal phase. (c,d) Crystalline structure in the cold region with higher packing fraction and nematic-like (N) structure in the hot region. This is far from the usual nematic phase as the breakdown of layering reduces the packing fraction significantly but the nematic order parameter is much higher than that in the equilibrium nematic phase. \Correction{(e) Packing fractions in the segregated zone corresponding to the aforementioned activities. The phases in each zone are mentioned for each activity. Parameters are the same as mentioned in Fig. \ref{iso}.}}\label{crys}
	
\end{figure*}

In the case of an initial crystal configuration, we surprisingly found $ \chi_{c} $ to be very high ($ \chi_{c} = 3.50-4.00 $) compared to the values at liquid phases. The reason is that many hot particles are stuck in the cold zone, and these particles require a larger amount of energy to overcome the potential barrier for demixing. Another reason is that the relaxation in the crystal phase is very slow compared to that in liquid crystal phases. Therefore, a longer time may be required to phase separate at smaller activity. After phase separation, the layered structure in the hot zone starts to break-up into a nematic-like structure that appears to be a far from equilibrium nematic phase (Fig. \ref{crys}-c,d). The local ordering is much higher
compared to that in the usual nematic phase; however, the breakdown of the
layered structure causes a significant decrease in the packing
fraction: $S_{hot} = 0.92$ and $\eta_{hot} = 0.62 $ at $ \chi = 9.00 $. Another important point to note is that, while melting, the hot particles do not go through a smectic phase. This differentiates the melting transition in the active subsystem  from the equilibrium one.

\subsection{Pair correlation functions} 

The local ordering in segregated zones are further characterized by calculating relevant pair correlation functions \cite{McGrother,bolhuis}. Apart from the radial distribution function $ g(r) $, we also calculate orientational pair correlation function $ g_{2}(r) $ which is relevant for quantifying nematic order. $ g_{2}(r) $ is defined as the $ 2^{nd} $ order Legendre polynomial associated with the orientation vectors $\vec{u_{i}}$ and $\vec{u_{j}}$ of two spherocylinders $i$ and $j$ separated by distance $r$: 
$ g_{2}(r) = \langle P_{2}(\vec{u_{i}}.\vec{u_{j}})\rangle$. We further calculate the 
vectorial pair correlation functions $ g_{\parallel}(r), g_{\perp}(r) $ which are the projections of the radial distribution function $ g(r) $ along the directions parallel and perpendicular to the nematic director, respectively. Periodic oscillations in $ g_{\parallel}(r) $ indicates the presence of layering and thus differentiate between nematic and smectic phases. $ g_{\perp}(r) $ indicates the presence of in-layer periodicity and thus differentiates between smectic and crystal phases.

\begin{figure*} [!htb]
	\centering
	\includegraphics[scale=1.30]{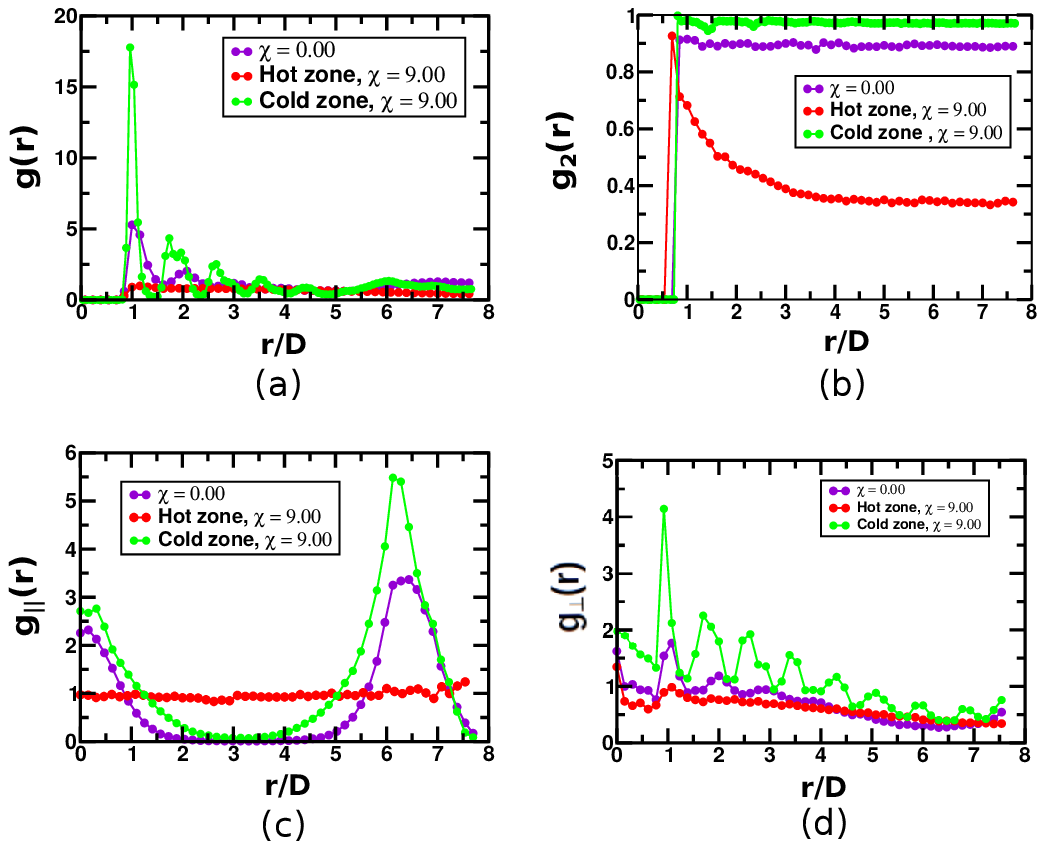}
	\caption{ Pair correlation functions in the segregated zones for a smectic initial phase ($ \eta = 0.67 $) at $ \chi = 9.00 $: 
		(a) center of mass pair radial distribution function $ g(r) $ (b) orientational pair radial distribution function $ g_{2}(r) $ (c)  projection of $ g(r) $ for the distances parallel ($ g_{\parallel} (r) $) and (d) perpendicular ($ g_{\perp} (r) $)  to the director of the spherocylinders.
	} \label{g-5}
	
\end{figure*}

In Fig. \ref{g-5}, we plot the pair correlation functions for a system starting from a smectic phase at $\eta = 0.67$ and $ \chi = 9.00 $. \Correction{The correlation functions are calculated in the hot and cold zone separately over a sphere of diameter 16D to quantify both short and long range correlations.} As shown in section (III-B-3), this system shows crystalline structure ($ S = 0.99 $) in the cold zone and nematic structure in the hot zone ($ S = 0.63 $) at this activity. From the pair correlation function $ g(r)$ shown in Fig. \ref{g-5}-a, we observe significant increase of the height of the $1^{st}$ and $2^{nd}$ peaks in the cold zone compared to the non-active case and the emergence of a $3^{rd}$ peak. This is a signature of high positional correlation among the cold particles. On the contrary, in the hot zone, we observe that the height of the 1st peak decreases significantly and the 2nd peak vanishes. However, orientational correlations (Fig. \ref{g-5}-b) still exists which identifies the phase as nematic. Periodic oscillations in $ g_{\parallel}(r) $ indicate the presence of a layered structure in the cold zone. The distance between two successive peaks is around 6.00 which is the end to end distance of a spherocylinder [$ L/D+1 $]. Multiple peaks at equal distance in $ g_{\perp}(r) $ signifies high translational ordering within the layer which confirms the emergence of local crystalline structure in cold particles' cluster.

\subsection{ Interfacial properties}

To obtain interfacial properties, we divide the simulation box into a number of slabs ($ N_{slabs}$). The number of slabs is chosen such that each slab contains enough particles (in our case, about $50 $ ) to get stable statistics. Effective density and temperature of the $ i^{th} $ slab are calculated as:

\begin{equation}
	\eta(i) =  \frac{n(i)}{v(i)}v_{hsc}
\end{equation}   

\begin{equation}
	5\times\dfrac{1}{2}k_{B}T_{eff}(i) = \dfrac{1}{n(i)}\sum_{j=1}^{n(i)} \left ( \dfrac{1}{2}mv_{j}^{2} + \dfrac{1}{2}I\omega_{j}^{2}\right ) 
	\label{eq-6}
\end{equation}   

\begin{figure} [!htb]
	\centering
	\includegraphics[scale=0.6]{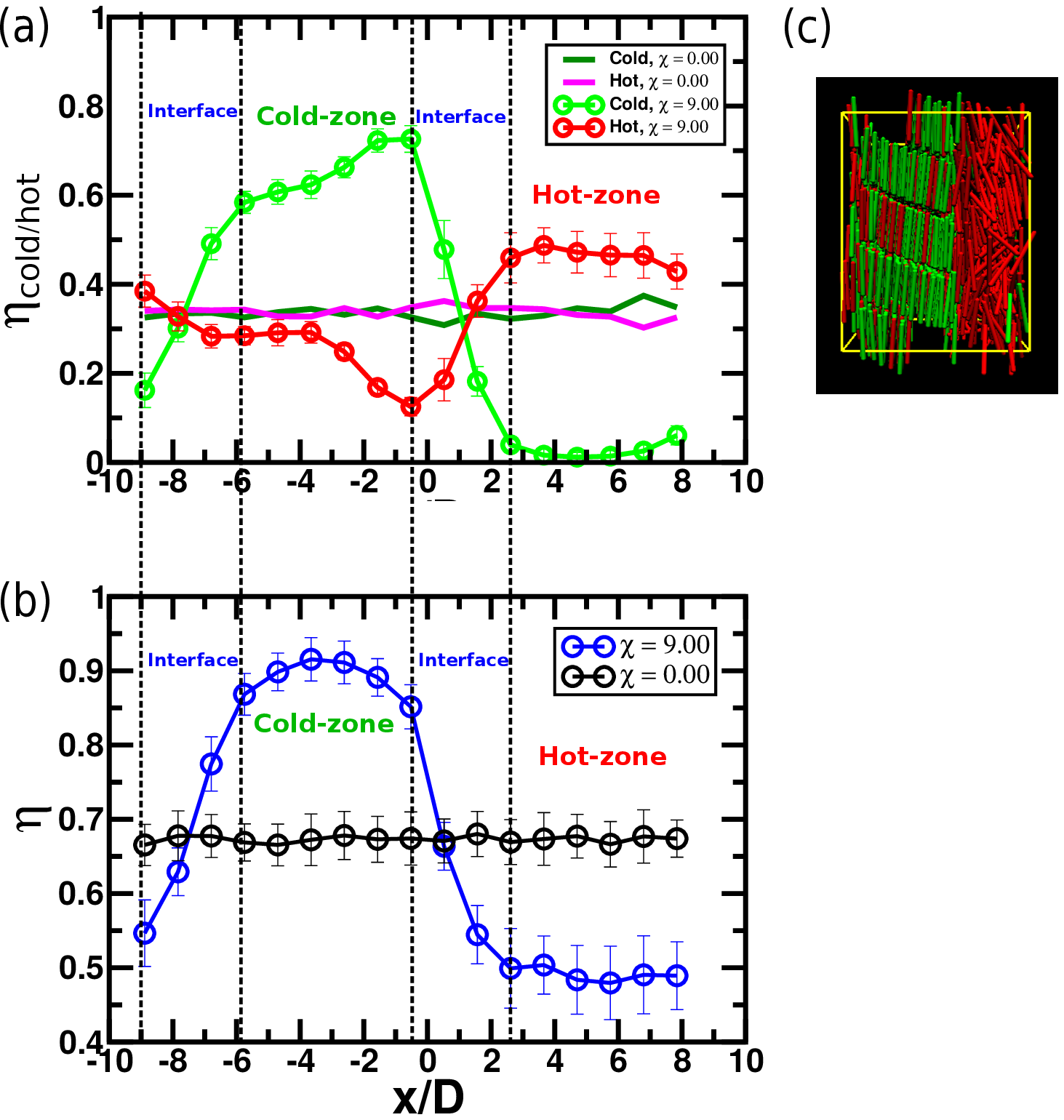}
	\caption{\Correction{(a) Packing fraction of cold $ \eta_{cold} $ and hot $ \eta_{hot} $ particles separately and (b) effective packing fraction of the particles present in each slab (including both hot and cold) along the direction perpendicular to the interface at $\eta = 0.67$, $\chi = 9.00$. Location of the cold, hot and interfacial regions are shown in the legend. The dotted lines indicate boundary of each zone. (c) Snapshot of the system at steady state.}} \label{A-interface}
	
\end{figure}

\begin{figure*} [!htb]
	\centering
	\includegraphics[scale=1.1]{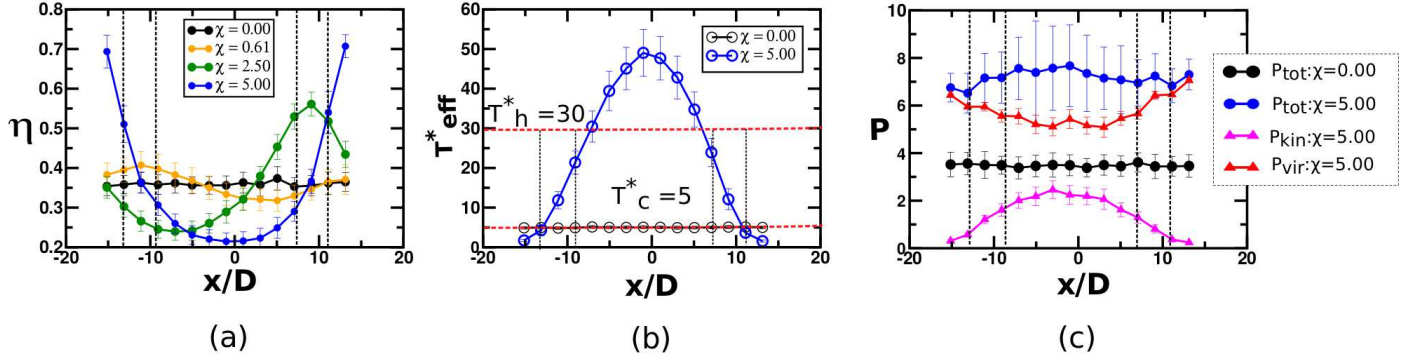}
	\caption{(a) Effective packing fraction $ \eta $, (b) temperature $ T^{*}_{eff} $ and (c) pressure $P$ profile along the direction perpendicular to the interface for different activities $ \chi $ at $ \eta = 0.36 $. The region between vertical dashed lines (black) indicate the location of the interface for $ \chi  = 5 $. Red horizontal dashed lines in Fig. (b) indicate the imposed temperatures on hot and cold particles at $ \chi = 5.00 $. In Fig.(c), we decompose the total pressure of the active system into kinetic and virial parts. It is clearly visible that the kinetic pressure is decreasing and the virial pressure is increasing from the hot to the cold zone. \Correction{0 is the position of the slab located at the half of the box-length along the perpendicular direction of the interfacial plane}.} \label{int-prop}
	
\end{figure*}

Here, $n(i)$ and $v(i)$ are the number of particles and volume of the $ i^{th} $ slab, respectively. $v_{j}$ and  $\omega_{j}$ indicate translational and rotational velocity of the  SRS $ j $, respectively. In equation (\ref{eq-6}), the term 5 appears on the left-hand side as the total number of degrees of freedom for a rigid spherocylinder is 5 arising from 3 translational and 2 rotational motions.
\Correction{We identify the locations of the phase-separated zones and the interface by calculating local packing fractions of hot ($\eta_{hot}$) and cold particles ($\eta_{cold}$) as shown in \Correction{Fig. \ref{A-interface}-(a)}. 
We observe the effective packing fraction of each slab $\eta$ (including both hot and cold particles) decreases in the hot zone and increases in the cold zone compared to the initial equilibrium system \Correction{(Fig. \ref{A-interface}-(b))}. The region where $\eta$ changes sharply from one zone to another is defined as the interface.} Our system exhibits two interfaces due to the effect of periodic boundary condition. The interface occupies a finite region of the simulation box and the width of the interface does not show any significant dependence on the system size, as shown in Fig. A5. In Fig. \ref{int-prop}-(a), we plot effective $\eta$ at different activities and find that the spatial inhomogeneity in $\eta$ increases with the increase of activity.


The effective temperature decreases continuously from hot to cold zone (Fig. \ref{int-prop}-b). The coexistence of two temperatures at the hot-cold interface signifies the non-equilibrium nature of the system. However, it is surprising that in the steady state, the cold zone has regions where the local temperature is lower than the imposed cold particles' temperature ($ T_{c}^{*} $) and in some regions of the hot zone, local temperature is much higher than the imposed hot particles temperature ($ T_{h}^{*}$). In Fig. \ref{int-prop}-b, we see that the maximum temperature in the hot zone is around 50 which is much higher than $ T_{h}^{*} = 30$. In the cold zone, the minimum temperature is around 2 which is lower than $ T_{c}^{*} = 5$. Though the effective temperature averaged over all the hot particles, $ T_{h}^{eff}$, is lower than $ T_{h}^{*}$ and the effective temperature averaged over all the cold particles, $ T_{c}^{eff}$, is higher than $ T_{c}^{*}$ due to heat exchange between them as mentioned in Eq-\ref{eq2}.

\begin{figure*} [!htb]
	\centering
	\includegraphics[scale=1.0]{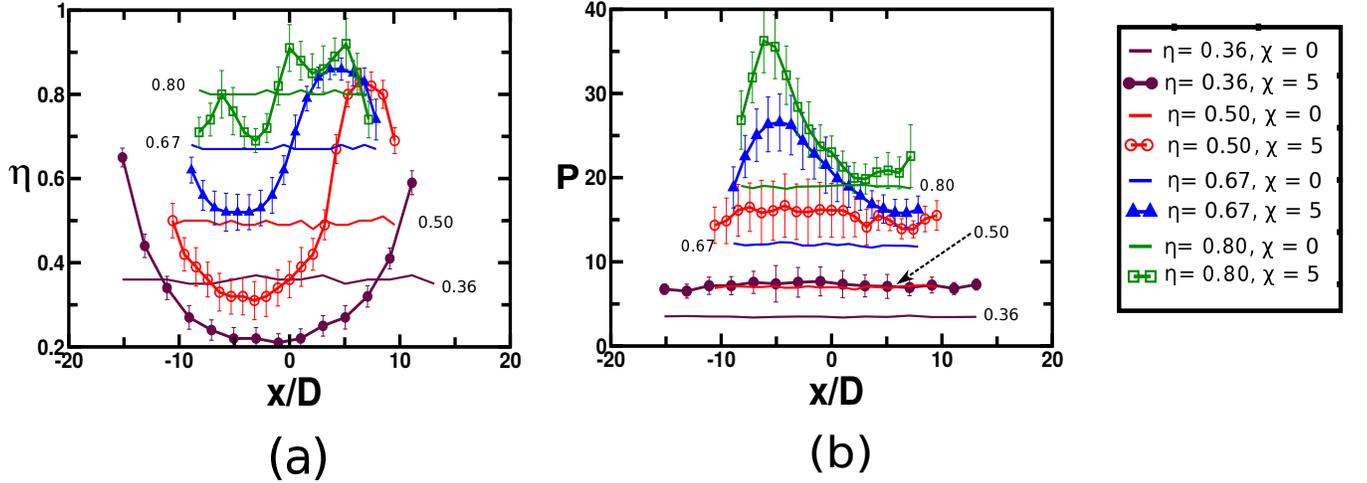}
	\caption{(a) Effective packing fraction ($ \eta $) and (b) pressure ($ P $) profile along the direction perpendicular to the interface for $ \chi = 0.00$ and $ 5.00 $ at different densities corresponding to the different initial phases: $ \eta = 0.36 $, initial isotropic (I) phase; $ \eta = 0.50 $, initial nematic (N) phase; $ \eta = 0.67 $, initial smectic (Sm) phase; $ \eta = 0.80 $, initial crystal (K) phase. The pressure is roughly constant within error bars for lower density phases (I, N) but it decreases continuously from the hot to the cold zone for higher density phases (SmA, K). \Correction{The equilibrium packing fraction corresponding to each solid line is mentioned.} \Correction{0 is the position of the slab located at the half of the box-length along the perpendicular direction of the interfacial plane}.}
    \label{int-pres}
	
\end{figure*}

We evaluate the pressure profile from diagonal components of the stress tensor. 

\begin{equation}
P_{kin}(i) = \frac{1}{3\times V(i)}\sum_{j=1}^{n(i)}  mv_{j}^{2} 
\end{equation}

\begin{equation}
P_{vir}(i) = \frac{1}{3\times V(i)}\sum_{j=1}^{n(i)-1} \sum_{k > j}^{}  \vec{r_{jk}}.\vec{f_{jk}}
\end{equation}

\begin{equation}
P(i) = \langle P_{kin}(i) + P_{vir}(i)\rangle_{ss}
\end{equation}
Here, $ P $, $ P_{kin}$, $ P_{vir} $ designate total, kinetic and virial pressures respectively. $ P_{vir} $ arises due to the particles' interaction which is defined as the product of relative distance $\vec{r_{jk}}$ and interacting force  $\vec{f_{jk}}$ between the SRS $j$ and $k$. We observe that, the local pressure increases with activity and it is nearly constant within error bars across the hot-cold interface (Fig. \ref{int-prop}-c). This is due to the fact that higher temperature causes higher kinetic pressure in the hot zone that acts at the hot-cold interface inducing higher packing and ordering in the cold zone. This enhances the virial pressure in the cold zone, resulting in the total pressure being constant across the interface. This is shown in figure \ref{int-prop}-(c) where we decompose the total pressure into kinetic and virial part for the system with $ \eta = 0.36 $, $ \chi = 5.00 $. However, this behavior is found only for the lower densities (initial phases I, N). In the case of higher densities (initial phases Sm, K), we observe the total pressure decreases continuously from the hot to the cold zone \Correction{(Fig. \ref{int-pres})}. This can be rationalized as follows: In the case of smectic and crystal initial phases, along with the kinetic pressure, virial pressure is also high in the hot zone due to their high orientational order ( as the hot zone shows nematic-like structure [Fig. \ref{sm}, \ref{crys}]). As a result total pressure increases in the hot zone which can not be compensated in the cold zone by increasing the virial pressure only. 
We calculate the pressure anisotropy ($ A(x) $) which is defined as:

\begin{equation}
	A(x) = P_{n}(x)- P_{t}(x)
\end{equation}

\begin{figure*} [!htb]
	\centering
	\includegraphics[scale=1.0]{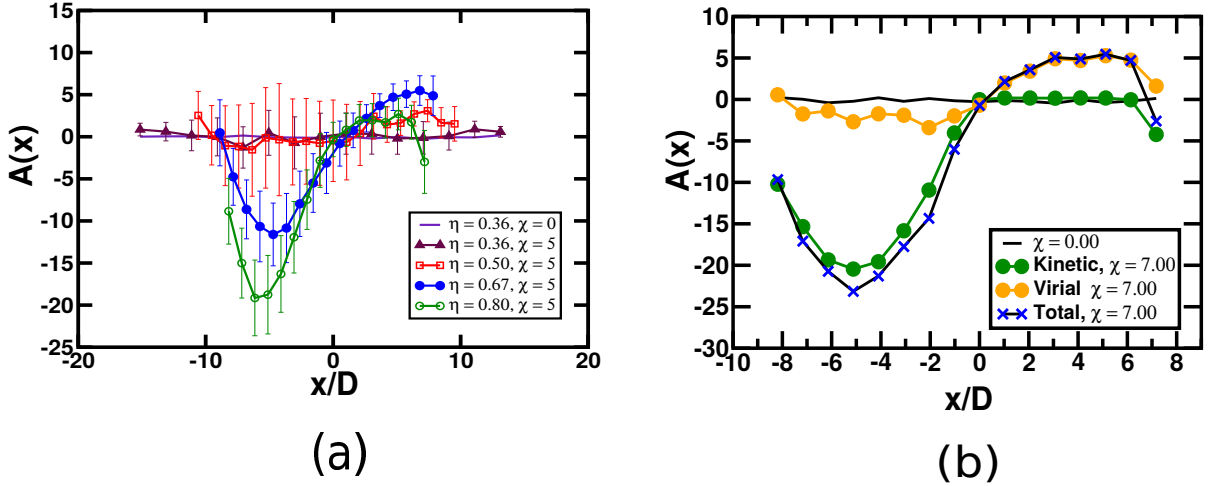}
	\caption{(a) Pressure anisotropy $ A $ as a function of distance $ x $ along the direction perpendicular to the interface for different packing fractions at $ \chi = 0.00 , 5.00 $. (b) $ A(x) $ in initial crystal phase ($\eta = 0.80$) for $ \chi = 7.00 $. Here, $ A(x) $ is decomposed into kinetic and virial parts. The pressure anisotropy is dominated by the kinetic part in the hot zone and virial part in the cold zone. \Correction{0 is the position of the slab located at the half of the box-length along the perpendicular direction of the interfacial plane}.} \label{int-anapres} 
	
\end{figure*}

\begin{figure*} [!htb]
	\centering
	\includegraphics[scale=0.9]{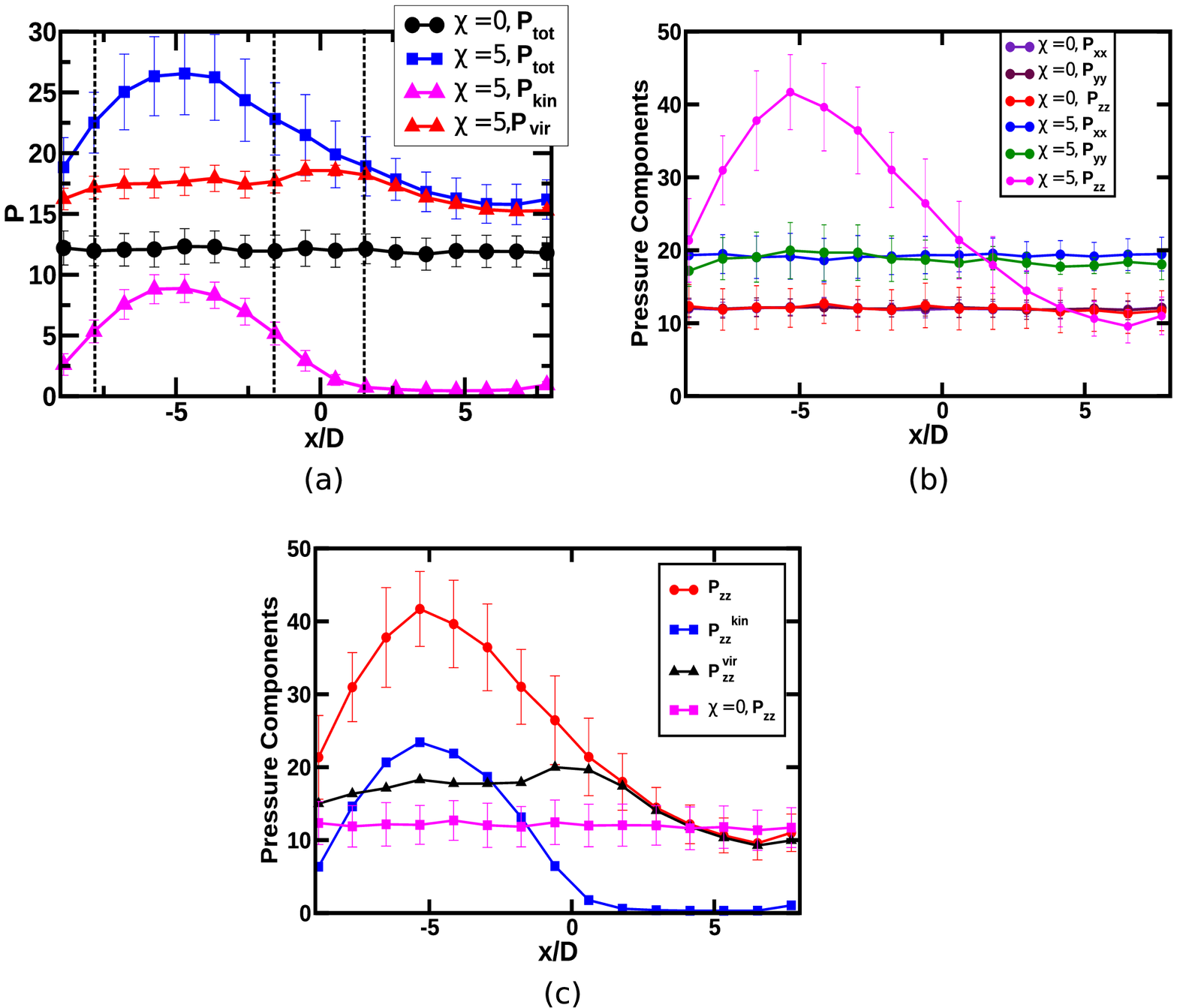}
	\caption{\Correction{(a) Pressure profile for $\eta = 0.67 $ at $ \chi = 0.00 , 5.00 $ along the direction perpendicular to the interface. Here, total pressure decreases continuously from hot to cold zone. The anisotropy in pressure is dominated by the kinetic contribution in the hot zone and the virial contribution (arising from wca interaction) in the cold zone. The region between the dotted lines indicate the interface. (b) Spatial variation of pressure components across the interface. Here $P_{xx}$ is the normal component and $P_{yy}$, $P_{zz}$ are the tangential components. The anisotropy of pressure is coming from $P_{zz}$ component which is also the nematic director of the hot zone. (c) $P_{zz}$ is decomposed into kinetic and virial part.}} \label{A-int-anapres-sm} 
	\end{figure*}

Here, $P_{n}(x)$ and $P_{t}(x)$ are the normal and tangential components of the total pressure respectively along the direction perpendicular and parallel to the interface. We designate the perpendicular direction of the hot-cold interface as   $ x $ and the other 2 directions parallel to the interfacial plane as $ y $ and $ z $. Thus the pressure components are defined as: $ P_{n}(x) = P_{xx}(x)$ and $ P_{t}(x) = (P_{yy}(x) + P_{zz}(x))/2 $. In equilibrium, pressure is isotropic, and therefore, $ A(x)=0 $. In the active system, we observe that the anisotropy is close to zero within the error bars for lower densities (initial phases I, N). However, it increases for higher densities (initial phases Sm, K) as shown in Fig. \ref{int-anapres}-(a).
To understand this precisely, we plot different pressure components across the interface as shown in \Correction{Fig. \ref{A-int-anapres-sm}} and find that the anisotropy in higher densities may arise due to the effect of active stress of the hot particles along their nematic director which acts parallel to the interfacial plane.
\Correction{To understand if the anisotropy in the pressure tensor is a consequence of the constant-volume simulation (NVT), we have done a constant-pressure simulation (NPT) with orthorhombic boundary condition and did not find a significant change in the pressure anisotropy. In references \cite{dominguez2002stress, bates1999computer,hashim1995computer}, it is reported that for equilibrium NPT simulation, the diagonal components of the pressure tensor become unequal in spatially ordered phases (smectic and crystal) due to maintaining a constant cubic shape throughout the simulation. In our case, NPT simulation is done in orthorhombic boundary conditions; hence this issue is not applicable in our case.}

\Correction{To understand possible system-size effects, we have done similar analysis with $N = 4096$ particles keeping all other system parameters and simulation protocol unchanged and we have not found any significant system size effect. The different results for two system sizes are compared in figure \ref{A-2}, \ref{A-3} and \ref{A-4} in the appendix.}

\section{\label{sec:level4} Conclusions and Future outlook}

In summary, we have presented a simple two-temperature model to study the thermodynamic and structural properties of active-passive spherocylinders where the level of activity is modeled by maintaining a temperature difference between the \textit{`hot'} and \textit{`cold'} particles. Starting from different initial equilibrium phases (I, N, Sm), we show that our simple model leads to not only phase separation into hot and cold regions but also liquid crystal ordering of the cold particles, and opposite shifts of the phase boundaries for mesophase formation, with respect to the equilibrium case, in the cold and hot domains. The extent of phase separation is quantified by an order parameter based on the local density. We find that the critical activity for phase separation lies in a small range, $ 1 < \chi_{c} < 4 $, for a wide range of densities from the isotropic to the crystal phase. This interesting observation highlights two-temperature model as an experimentally feasible system for studying effect of scalar activity in colloidal rods. We observe that the critical activity decreases with density in the liquid regime and increases again in the crystal regime. Based on these observations, a phase diagram is drawn in the state phase, $ \chi $ vs. $ \eta $, showing the parametric regions of phase-separated and homogeneously mixed states.

We find the segregated zones developing different liquid crystal structures depending on the activity and initial phase of the system. For example, an initial isotropic configuration shows nematic ordering in the cold region, which eventually turns into crystalline ordering at higher activities. Similarly, an initial nematic configuration shows smectic or crystal ordering in the cold zone, depending on the value of $\chi$, and isotropic structure in the hot zone. As a result, the I-N phase boundary shifts towards higher density for the hot particles and lower density for the cold particles. The segregated structures are identified by calculating the local nematic order parameter and different pair correlation functions. Finally, we analyse interfacial profiles of various thermodynamic quantities and conclude that the order-disorder transitions in the segregated zones are probably governed by local balance of pressure across the interface: higher temperature induces higher kinetic pressure in the hot zone which is compensated in the cold zone by an increased virial  pressure. 

Another possible reason for the order-disorder transition may be an entropic effect. The hot particles compensate for the loss of entropy due to ordering transition in the cold zone by developing a disordered structure. An important component of our future work will be to examine configurational entropy and free energy, as well as entropy production and currents, to shed light on the mechanisms underlying the nonequilibrium phase transitions we observe. 
Finally, analytical theories of two-temperature models are so far limited to spherical particles \cite{joanny-2015,joanny-2018,joanny-2020}. Hence, 
generalizing their theoretical approach to make analytical predictions for  two-temperature models with anisotropic particles is an important challenge.


\begin{acknowledgments}
We would like to thank Aparna Baskaran for helpful discussions. We also thank Prof. Yves Lansac for insightful suggestion. We thank DAE, India for financial support through providing computational facility. JC acknowledges support through an INSPIRE fellowship. SR was supported by a J C Bose Fellowship of the SERB, India, and by the Tata Education and Development Trust, and acknowledges discussions during the KITP 2020 online program on Symmetry, Thermodynamics and Topology in Active Matter. This research was supported in part by the National Science Foundation under Grant No. NSF PHY-1748958. CD was supported by a Distinguished Fellowship of the SERB, India.
\end{acknowledgments}

\bibliography{ref}

\appendix*


\Correction{\section{Quantifying macroscopic phase separation:}
To quantify if the phase-separation happens in the macroscopic level, we have used the following critera:
we divide the simulation box into a number of slabs ($ N_{slabs} $) along the direction normal to the interface. $ N_{slabs} $ is chosen such that each slab contains enough particles (in our case, about $50 $ ) to get stable statistics. For each slab $ i $, we calculate the number difference of hot ($ n_{h}^{i} $) and cold ($ n_{c}^{i} $) particles  divided by total number of particles ($ n_{tot}^{i} $) in that slab. We then define:}

\Correction{\begin{equation}
	\phi_{i} = \left\langle \frac{n_{c}^{i}-n_{h}^{i}}{n_{c}^{i}+n_{h}^{i}} \right\rangle _{ss}
\end{equation}}

\begin{figure*} [!htb]
	\centering
	\includegraphics[scale=1]{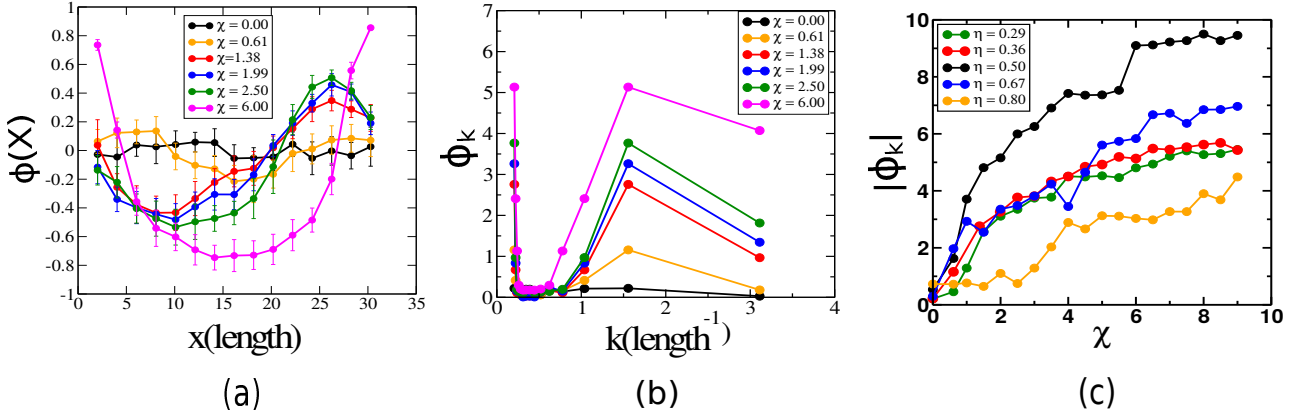}
	\caption{(a) Density order parameter $ \phi $ at $ \eta = 0.36 $ for different activities $ \chi $ along the direction normal to the interface. (b) Fourier components of $ \phi(x) $. Here, $ k = 2\pi/x$. In this figure we can see 2 peaks in $ \phi_{k} $: the peak at lowest k value (around $k = 0.2$ $\Rightarrow$ $x = 31.42$ which is approximately the length of the box L)  indicates occurrence of phase separation in macroscopic level and the other peak (around $k = 1.5 $ $\Rightarrow$ $x = 4.2 $) indicates microscopic phase separation. The amount of phase separation is determined from the height of the 1st peak in $ \phi_{k} $. \Correction{(c) Magnitude of the Fourier component of $ \phi $ for the smallest $k$ vs activity $ \chi $ at several packing fractions ($ \eta $). It shows similar trend as shown in Fig. \ref{denop} in the main text.}}\label{A-1}
\end{figure*}

\Correction{where $\langle \dots \rangle_{ss}$ represents a steady-state average over a sufficiently large number of configurations. The average of $\phi_i$ over all slabs will be 0 as $\phi_i$ varies symmetrically from $-1$ to $+1$ from the hot-rich to the cold-rich zone. We therefore calculate Fourier transformation of $\phi(x)$ where $x$ measures the position of the slab. The magnitude of the first non-vanishing Fourier component $|\phi_{k}|$ is the measure of macroscopic phase separation in our system (Fig. \ref{A-1}). In Fig. \ref{A-1}-b, the 1st peak occurs at x = 31.42 which is approximately the length of our simulation box (L = 32). This indicates occurrence of phase separation in macroscopic scale. In Fig. \ref{A-1}-c, we plot the magnitude of $|\phi_{k}|$ for the smallest $k$ as a function of the  activity at several packing fractions that shows similar trend as shown in Fig. \ref{denop} in the main text. }


\centerline{}

\begin{figure*} [!htb]
	\centering
	\includegraphics[scale=1.0]{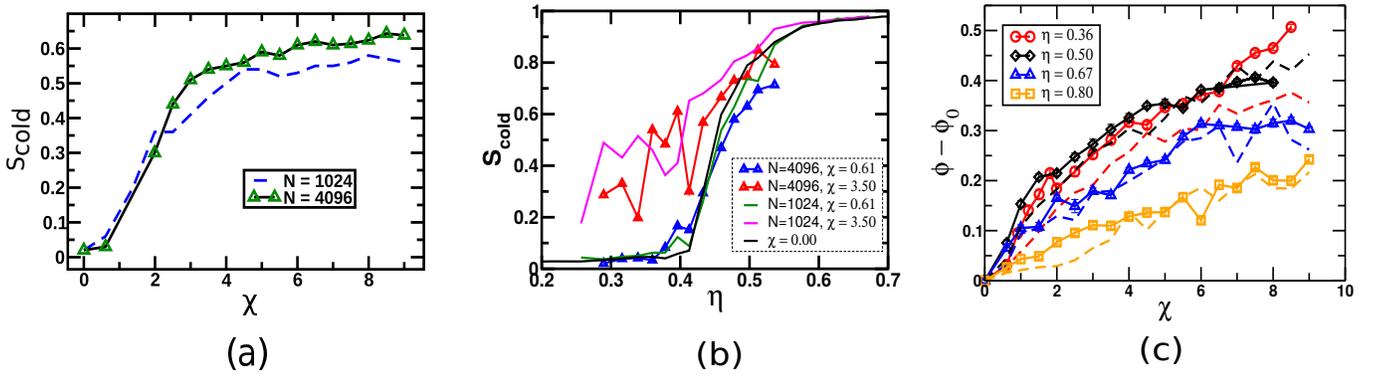}
	\caption{(a) Average nematic order parameter of cold particles $S_{cold}$ with activity $\chi$ for $N = 4096$ (solid lines) compared to the earlier case with $N = 1024$ (dashed line) at $ \eta = 0.36 $. (b) $S_{cold}$ vs $ \eta $ at different $\chi$ as mentioned in the legend. Here we also observe local minima for a certain range of $\eta $ as it is observed in the earlier case with $N = 1024$ Fig.-\ref{hot-cold-5}-(a). \Correction{(c) Density order parameter ($\phi$ ) with $\chi$ for the larger system size with $N = 4096$ at different packing fractions. The density order parameter in the system with $N = 1024$ particles are designated by dotted lines at the respective packing fractions. We have not observe any significant system size effect for these cases.}} \label{A-2}
	
\end{figure*}


\begin{figure*} [!htb]
	\centering
	\includegraphics[scale=1.70]{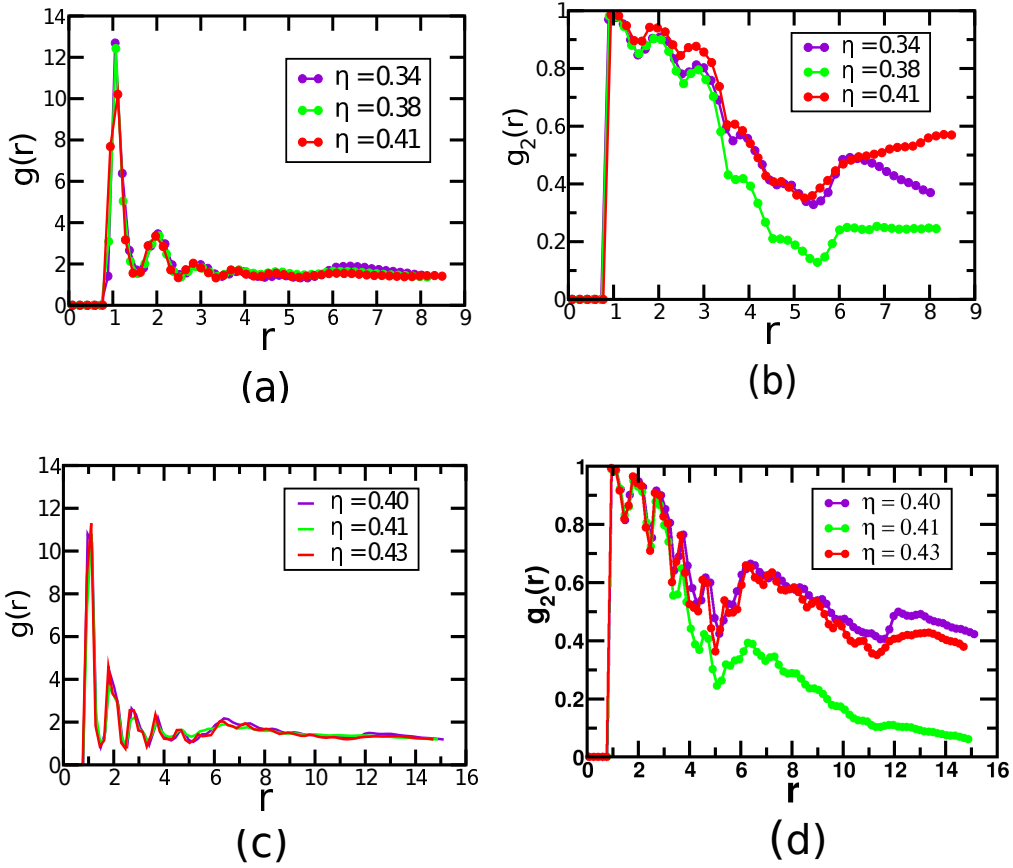}
	\caption{Pair correlation functions for $\chi = 3.50 $ at the values of $ \eta $ around which we find local minima in $ S_{cold} $ for $N=1024$ and $N=4096$ as mentioned in Fig. \ref{A-2}-(b). The local minima occurs at the value of $ \eta = 0.38 $ and $ \eta = 0.41 $ for $ N = 1024 $ and $ N = 4096 $ respectively. For both of the cases, orientational correlation $ g_{2}(r) $ decreases and saturates at a smaller value at large distances (panels (b) and (d)). This is due to the presence of local domains of different average orientations of the nematic director which effectively decreases the global ordering of the cold particles. For other packing fractions, $ g_{2}(r) $ saturates at a higher value at large distances. Here we observe a stable single domain of a fixed orientation of the nematic director in the cold zone. Though the translational correlation $g(r)$ (panels (a) and (c)) is nearly independent of $\eta$ in the range considered.} \label{A-3}
	
\end{figure*}

\begin{figure*} [!htb]
	\centering
	\includegraphics[scale=1.0]{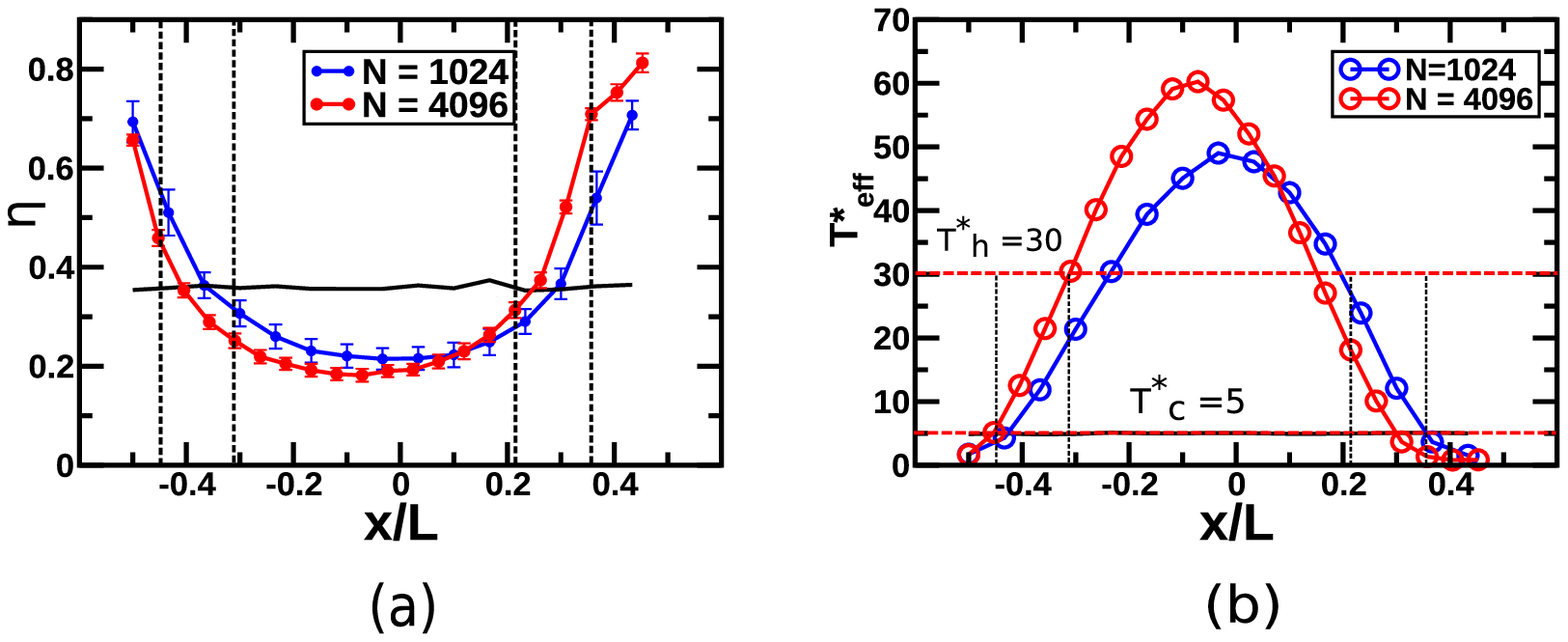}
	\caption{(a) Effective packing fraction $ \eta $ and (b) temperature $ T^{*}_{eff} $ profile along the direction perpendicular to the interface scaled by the box-size for $N = 4096 $ and 1024 at $ \eta = 0.36 $, $ \chi = 5.00 $. The black horizontal solid line in (a) indicates the local packing fraction in absence of activity.
	Black vertical dotted lines in (a) and (b) indicate the interfacial region. Red horizontal dashed lines in (b) indicate the imposed temperatures on hot and cold particles. For both system sizes, the width of the interfacial region is about 5 in units of the diameter (D) of the SRS.} \label{A-4}
	
\end{figure*}


\end{document}